\documentclass{PoS}
\def\mpi2{m_\pi^2}
\def\mK2{m_K^2}

\newcommand{\bea}{\begin{eqnarray}}
\newcommand{\eea}{\end{eqnarray}}
\newcommand{\be}{\begin{equation}}
\newcommand{\ee}{\end{equation}}

\newcommand{\VEV}[1]{\left\langle #1\right\rangle}

\newcommand{\Tr}{\mbox{Tr}}

\newsavebox{\DERIVBOXZLM}
\savebox{\DERIVBOXZLM}[2.5em]{$\Longrightarrow\hspace{-1.5em}
\raisebox{.2ex}{*}
\hspace{-.7em}\raisebox{-.8ex}{\scriptsize lm}\hspace{.7em}$}

\usepackage{fleqn,epsf}
\title{Effects of the disconnected flavor singlet corrections on the hyperfine
splitting in charmonium}

\ShortTitle{Effects of the disconnected flavor singlet corrections on the hyperfine
splitting in charmonium}

\author{
         C. DeTar and \speaker{L. Levkova}\\
        University of Utah, Salt Lake City, UT 84112, USA\\
        E-mail: \email{detar@physics.utah.edu, ludmila@physics.utah.edu}}
\author{Fermilab Lattice and MILC Collaborations}
\abstract{Experimentally the charmonium hyperfine splitting 
is $M_{J/\psi}- M_{\eta_c}=117$ MeV and current lattice results 
are generally below this value. The difference could be due to 
the effects of the disconnected flavor singlet diagrams which 
have not been included in these calculations. Previous attempts
 to determine the disconnected flavor singlet corrections have 
led just to rough estimates in the range of $\pm 20$ MeV. We present 
preliminary results for these corrections calculated on fine 
($a\approx 0.09$~fm) Asqtad 2+1 flavor lattices provided by the
MILC Collaboration.}

\FullConference{The XXV International Symposium on Lattice Field Theory\\
                 July 30 - August 4 2007\\
                 Regensburg, Germany}

\begin{document}

\section{Introduction}
The lattice calculation of the hyperfine splitting in charmonium is
currently still showing a discrepancy with the experimental value of
117 MeV. The discrepancy is large (30-40\%) in the quenched case \cite{qu1,qu2}.
It is reduced to around 10\% in dynamical studies with improved actions
\cite{on,hisq}, 
but its origin is still uncertain. A possible
explanation could be that even the current state-of-the-art lattice
action formulations do not reproduce the quark dynamics within the charmonium
states sufficiently accurately for the purpose of this calculation. 
Another possibility which could account for the discrepancy (or at least 
part of it) is the neglected contribution of the disconnected diagrams 
in the lattice computations. These diagrams
could contribute to the masses of both the vector $J/\psi$ 
and the pseudoscalar $\eta_c$, and thus affect the hyperfine splitting
$M_{J/\psi}-M_{\eta_c}$. Perturbatively, the
contribution of these diagrams  in charmonium is expected to be small
due to the OZI suppression, especially for the vector state.
However, non-perturbative effects, such as the $U_A(1)$ anomaly \cite{u1}
and glueball mixing,
 might enhance it
enough so that it becomes a non-negligible fraction of the hyperfine splitting.

In this work, we present our preliminary results of the effort to test this
possibility and determine the contributions of the disconnected diagrams.
Previous calculations \cite{fs1,fs2} using two-flavor gauge ensembles very 
roughly estimate
the contribution to be within $\pm20$ MeV. They both confirm that there are 
significant difficulties in obtaining a signal for the disconnected diagrams
due to noise, especially for heavy quarks. 
Our calculations are performed on an ensemble of 505, 2+1 flavor
 MILC lattices \cite{Bernard:2005ei}, generated with the improved Asqtad action \cite{asq}. For this ensemble 
the ratio of the masses of the light and heavy 
quarks is $m_{ud}/m_s=0.1$.
The charm quarks are simulated using the Fermilab
interpretation of the clover action with $\kappa_c=0.127$ tuned to
the physical charm quark mass. The disconnected diagrams are calculated stochasticly
 with 12 spin and color diluted sources. 

We employ several methods for 
improving our calculation in comparison with previous works. First, 
our lattice volume is $40^3\times 96$, which is much larger than in \cite{fs1,fs2}
and our lattice scale, $a\approx 0.09$ fm, is finer. In addition to
these improvements we use the unbiased subtraction technique \cite{unb}
in the stochastic estimators.
The success of this technique depends on the fast convergence
of the hopping parameter expansion used in the subtraction.
Considering that $\kappa_c$ is small for the charm quark, we use
the terms of the expansion only up to third order in $\kappa_c$, which reduces the
standard deviation of the disconnected correlator  by about a factor of 4. The last and 
most important improvement is that we study the 
point-to-point (PP) disconnected correlators instead of the traditional
time-slice-to-time-slice (TS) ones. As a result, the standard deviation of the signal
is reduced about three orders of magnitude.
  
\section{Lattice formulation}
The disconnected part of the flavor singlet TS correlator 
is calculated as:
\be
D(t)=c_\Gamma\VEV{L(0)L^\ast(t)},\hspace{0.5cm}{\rm where}\hspace{0.5cm}
L(t) = \Tr(\Gamma M^{-1})
\ee
and the trace is over the Dirac, color and space indices. For the vector
we have $\Gamma=\gamma_\mu$, $c_\Gamma=1$ and for the pseudoscalar $\Gamma=\gamma_5$,
$c_\Gamma=-1$. 
To determine the effect of the disconnected diagram on the masses of
the charmonium states,
previous works \cite{fs1,fs2} explore the ratio of the disconnected to the connected
TS correlators at zero momentum:
\be
\frac{D(t)}{C(t)} = \frac{F(t)}{C(t)} -1 =\frac{A_f}{A_c}\,e^{(m_c -m_{f})t} -1.
\ee
In the above $F(t)=D(t)+C(t)$ is the full propagator corresponding to a
state with ``full'' mass $m_f$. The mass $m_c$ extracted from the
connected propagator is the mass which is usually studied 
in lattice simulations. The constants $A_f$ and $A_c$ are  
the full and connected propagator amplitudes, respectively. 
Considering that the available lattices are quenched with respect to the
charm quark, an appropriate fitting form for the ratio data  at zero momentum would be
\be
\frac{D(t)}{C(t)} = (m_c -m_{f})t + \frac{m_c -m_{f}}{m_c}.
\ee
The differences $m_c -m_{f}$ should be calculated for both 
the vector and the pseudoscalar and then subtracted from the mass of each 
in order to determine the effects of the 
disconnected diagrams on the hyperfine splitting.

In this work however, we calculate the PP disconnected 
correlators, and the above analysis has to be modified accordingly.
The disconnected PP propagator as a function of the Euclidean distance 
on the lattice $r$ is defined as:
\be
D(r)=\frac{c_\Gamma}{N_r}\sum_{r=\left|x-x^\prime\right|}\VEV{L(x)L^\ast(x^\prime)},
\ee
where
the sum is over all pairs of lattice points at this distance,
$N_r$ is the number of these pairs and there is trace only over spins and colors
in $L$. The connected PP correlator
$C(r)$ is defined in a similar manner.
It is known that in the continuum limit the asymptotic behavior of $C(r)$
is:
\be
C(r)\sim A\frac{e^{-m_cr}}{r^{\frac{3}{2}}},
\ee
and consequently the behavior of $D(r)$ in this case can be deduced as follows:
\be
D(r)\sim -\frac{d}{dm_c^2}C(r) \sim B\frac{e^{-m_cr}}{r^{\frac{1}{2}}}.
\ee
Thus their ratio is
\be
\frac{D(r)}{C(r)} \approx \frac{B}{A}r,
\ee
where we interpret the amplitude ratio as:
\be
\frac{B}{A} = m_c-m_f.
\ee
The behavior of $D(r)$ and $C(r)$ on the lattice will be
affected by discretization artifacts; still, for large $r$
and fine lattices we assume that the above behavior is a good 
approximation. 
%Thus the purpose of our calculation will be to determine
%the amplitudes $A$ and $B$ from $C(r)$ and $D(r)$ data by fitting
%to the the above forms. 

\section{Results for $D(r)$}
From the previous studies it is known that the charmonium correlator signal
disappears very quickly around $t=2-3$.
We work with the PP disconnected propagator, since this way we benefit
from both the additional data at non-integer distances 
and the much improved statistics.
The correlator $D(r)$ has from one to three orders of magnitude smaller 
relative errors than the TS
disconnected propagator in the region where we have a signal. Figure~1
illustrates this statement by comparing $D(r)$ and $D(t)$ for $\eta_c$
for two different ranges of $r$ and $t$. 
In the right panel of Fig.~1, the comparison is done on 
a shorter range in order to emphasize the fact that we do have a clear signal
for $D(r)$ in the range where the $D(t)$ signal is completely 
obscured by the noise. 
\begin{figure}[ht]
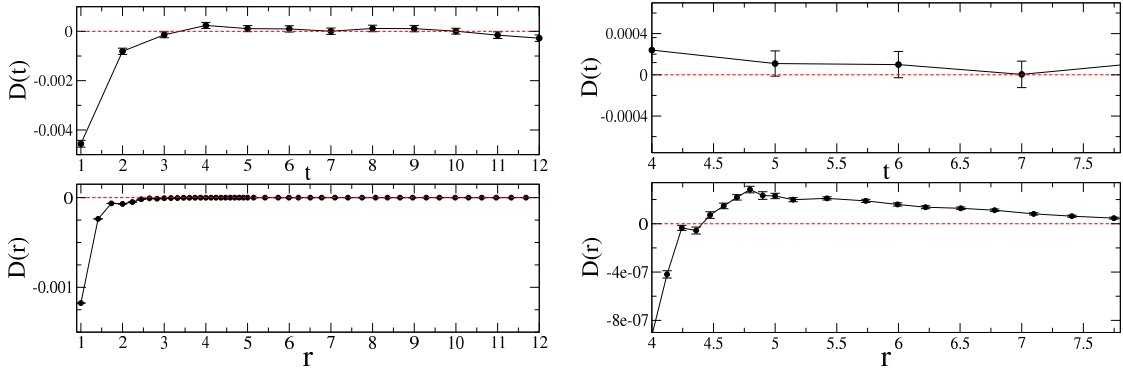

\begin{tabular}{cc}
 \epsfxsize=72mm
  \epsfbox{comparison.eps}
&
  \epsfxsize=72mm
  \epsfbox{comparison1.eps}
\end{tabular}
\caption{Comparison of $D(t)$ and $D(r)$ for $\eta_c$ 
for two different ranges of $t$ and $r$.}
\label{fig:PP_vs_TS}
\end{figure}
The result that the $D(r)$ signal is so much better than the one for $D(t)$ can
be explained by the fact that in the calculation of $D(t)$ there are a 
great number of contributions from points, which although not far from each other
in the $t$ direction, are far in the 4d Euclidean space.
For the disconnected correlator, the noise increases strongly with the distance
and such points contribute nothing to $D(t)$ but noise. This problem is
avoided when working with $D(r)$ instead.

\section{The $D(r)/C(r)$ ratio for $\eta_c$}
Figure~2 shows our data for the ratio $D(r)/C(r)$.
 \begin{figure}[ht]
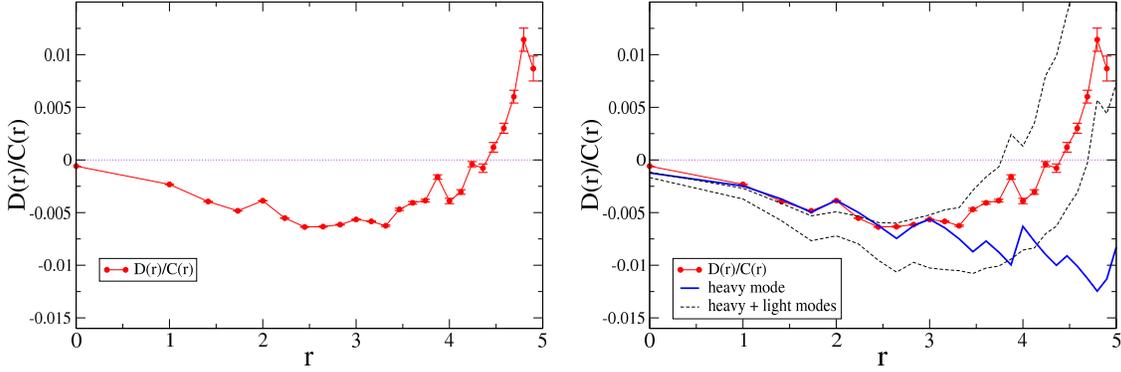

\begin{tabular}{cc}
 \epsfxsize=72mm
  \epsfbox{Results2aa.eps}
&
  \epsfxsize=72mm
  \epsfbox{Results.eps}
\end{tabular}
\caption{The lattice data for the ratio $D(r)/C(r)$ for $\eta_c$. In the right
panel the actual data is compared with theoretical models with a heavy mode only
or with a sum of heavy and light modes.}
\label{fig:ratio}
\end{figure}
We make the following observations:
\begin{itemize}
  \item The roughness of the data at short distance is a sign of
    rotational symmetry breaking and requires a nonasymptotic lattice
    fitting function.
  \item The ratio $D(r)/C(r)$ inherits the sign flip of $D(r)$.  The
    sign flip is expected, since from the CP symmetry of the operator,
    the sign of $D(r)$ at $r = 0$ must be opposite the sign of any
    single-pole contributions, and they dominate at large $r$. We believe this
    is the first observation of light state contributions to an
    operator constructed from charm quarks.
  \item In a separate analysis we find that excited states in the
    TS correlator $C(t)$ contribute more than 20\% for $r < 5$.
    The same is expected for $D(r)$.  Indeed, if we assume a single
    heavy mode at short distance, to make the nonasymptotic form of
    $D(r)/C(r)$ resemble the data, we need a mass much heavier than
    the $\eta_c$, as indicated in the right panel of
    Fig.~\ref{fig:ratio}.  The light states in $D(r)$ are very likely
    dominant for $r > 4$, but we estimate that the charmonium signal
    is still about 30\% of the whole signal at $r=5$.
  \item The asymptotic expression Eq.~(2.7) is not likely to be
    applicable anywhere. At large distance $D(r)$ is dominated by
    lighter hadronic modes (the $\eta$ and $\eta^\prime$ mesons,
    glueballs, etc.). Thus to extract the ratio of ground state
    amplitudes we must resort to more elaborate models.
  \item In the right panel of Fig.~\ref{fig:ratio} we present two more
    exploratory approximations for $D(r)/C(r)$ that include one heavy
    and one light mode, just to show that the behavior of the ratio
    data can be approximated in this manner.
\end{itemize}

\section{Extracting the ground state $\eta_c$ signal from $D(r)$}

Following the discussion in the previous section, at present we are
led to explore a simplified model for the intermediate range $r \in [5,11]$
that includes a ground and excited $\eta_c$ state and a light state.
The light state represents a number of possible states.  We work
separately with $D(r)$ and $C(r)$, extract the amplitudes of the
$\eta_c$ contributions, and finally use Eq~(2.8) to get an approximate
mass shift.

For the connected correlator we have only a connected-$\eta_c$ ground
state and one excited state. For the disconnected correlator our
ansatz includes a light state as well.  In momentum space it reads
\be
   D(p^2) = \left[\frac{f(p^2)}{p^2+m_c^2} + 
           \frac{g(p^2)}{p^2+m_c^{\ast 2}}\right]^2\frac{1}{p^2 + m_l^2}
\ee
where $m_l$ is a light state mass, $m_c^\ast$ is an excited connected
$\eta_c$ mass, and $f(p^2)$ and $g(p^2)$ are real functions.  We need
the residue of the double pole at $p^2 + m_c^2 = 0$ to determine the
mass shift.  In coordinate space the asymptotic form is then
\be
D^{fit}(r)= \frac{B}{r^{\frac{1}{2}}}(e^{-m_cr}+
e^{-m_c^\ast r}) + \frac{cB}{r^{\frac{3}{2}}}({e^{-m_cr}-e^{-m_c^\ast r}}) + \frac{L}{r^{\frac{3}{2}}}e^{-m_lr}.
\ee
\begin{figure}[ht]
\begin{tabular}{ccc}
 \epsfxsize=47mm
  \epsfbox{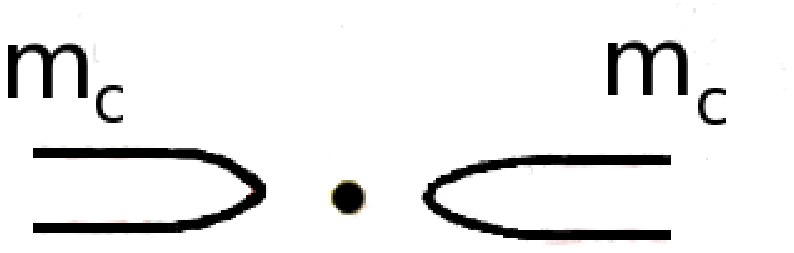}
&
  \epsfxsize=47mm
  \epsfbox{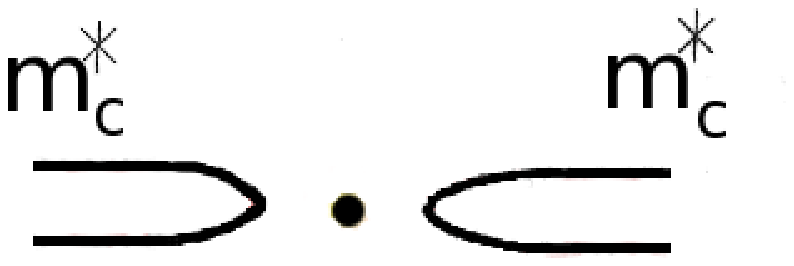}
&
  \epsfxsize=47mm
  \epsfbox{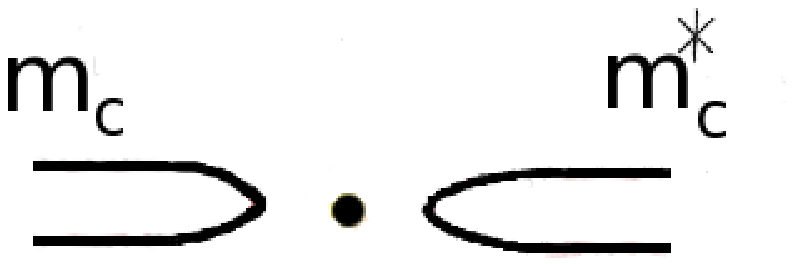}
\end{tabular}
\caption{Charmonium disconnected diagrams included in the fitting form Eq.~(5.2). The
ground state diagram is to the left, the excited is in the middle and the ``mixed''
one is to the right.}
\label{fig:diagram}
\end{figure}
The terms correspond, respectively, (Fig.~\ref{fig:diagram}) to double pole
contributions from the ground and excited charmonium states, mixed
ground-excited state contributions, and the light state contribution.
For this preliminary analysis, we take the amplitudes ($B$) of
the ground and excited charmonium states to be equal. We do this in order to
reduce the number of free parameters. Since the excited state
contribution is just a small correction in the chosen fit range, this
restriction on the amplitude is not of great importance.  The
coefficient $c$ is determined from the mass terms in the ansatz.  We
fix $c\approx 7$, but find that our results are not very sensitive to
varying it over the range $c\in[2,14]$.

We adjust only $B$ and $L$.  We fix the masses as follows: The light
mass is taken to be the central value $m_l=0.43(1)$ from a
single-exponential fit (as in Eq.~(2.5)) in the range $r \in [7,12]$.  The connected
$\eta_c$ and $\eta_c^\ast$ masses, $m_c=1.1598(7)$ and
$m_c^\ast=1.51(5)$, are known from fits to the connected TS propagator
$C(t)$.

\begin{figure}[ht]
 \epsfxsize=100mm
\begin{center}
  \epsfbox{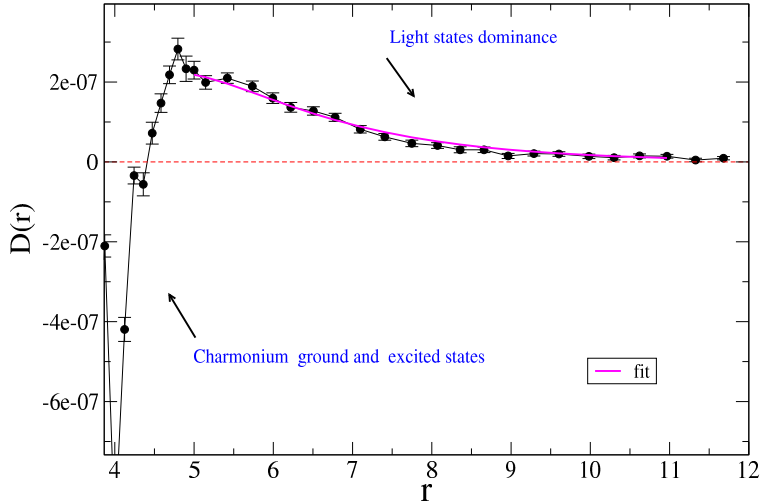}
\caption{The disconnected PP correlator $D(r)$ and fit to it in the
range of distances $r\in[5,11]$.
}
\label{fig:disc_fit}
\end{center}
\end{figure}

Before fitting we smoothed the data by averaging the signal in small
bins in $r$, thus reducing the effects of the lattice artifacts and
improving the continuum approximation.  The resulting fit to $D(r)$ is
shown in Fig.~\ref{fig:disc_fit}.  The fit has $\chi^2/df = 24 / 19$
.
Our fit result favors the following range of values for the ratio:
\be
\frac{B}{A} = m_c-m_f\in [-4, -1]\,\,\,{\rm MeV}.
\ee
Thus the disconnected diagram contribution slightly increases the
$\eta_c$ mass.

\section{Summary and conclusions}

We have calculated the disconnected pseudoscalar propagator with
dramatically reduced noise using unbiased subtraction methods and
point-to-point correlator data.  We fit the data to a
simplified model and extracted the amplitudes of the ground state in
the disconnected and connected propagators.  We find that the
disconnected diagram contribution increases the $\eta_c$ mass by 1-4
MeV.  

The vector meson $J/\psi$ can be studied in a similar fashion.  If the
corresponding mass shift of the $J/\psi$ is negligible as expected
\cite{fs1,fs2}, we would then conclude that the hyperfine splitting
is reduced slightly.

This result is preliminary, however.  Further work is needed to
explore the sensitivity of this conclusion to the choice of the
fitting model and to include effects of rotational symmetry breaking
properly.  Calculations at a smaller lattice spacing are needed to
test for heavy-quark and hard-gluon discretization errors.  A similar
study of the $J/\psi$ could test the common expectation that its mass
shift is negligible.

\section*{Acknowledgments}
We are grateful to Peter Lepage for helpful comments.
This work was supported by the US DOE and NSF.
Computations were performed at CHPC (Utah) and FNAL.

\end{document}